\documentclass[reprint,prfluids,onecolumn,longbibliography]{revtex4-2} % reprint 
\usepackage{graphicx}
\usepackage{subfigure}
\usepackage{amssymb}
\usepackage{amsfonts}

\usepackage{amsmath}
\usepackage{amsbsy}
\usepackage{mathrsfs} 
\usepackage{color}
\usepackage[colorlinks=true,citecolor=blue]{hyperref}
\usepackage{soul}%\st{...}
\usepackage{lineno}
\usepackage{mymacros}
\usepackage{graphicx}
\begin{document} 

%\title{Onset of drop Quincke rotation: bistability and charge-density blowup}
\title{Bistability and charge-density blowup in the onset of drop Quincke rotation}
\author{Gunnar G. Peng$^1$, Ory Schnitzer$^2$}
\affiliation{$^1\!$Department of Mathematics, University College London, London WC1E 6BT, UK\\
$^2\!$Department of Mathematics, Imperial College London, London SW7 2AZ, UK}

\begin{abstract}
Particles in a sufficiently strong electric field spontaneously rotate, provided that charge relaxation is slower in the particle than in the suspending fluid. It has long been known that drops also exhibit such ``Quincke rotation,'' with the electrohydrodynamic flow induced by electrical shear stresses at the interface leading to an increased critical field. However, the hysteretic onset of this instability, observed for sufficiently low-viscosity drops, has so far eluded theoretical understanding---including simulations that have struggled in this regime owing to charge-density-steepening effects driven by surface convection. Here, we conduct a numerical study of the leaky-dielectric model in a simplified two-dimensional setting involving a circular drop, considering arbitrary viscosity ratios and field strengths. As the viscosity of the drop is decreased relative to the suspending fluid, the pitchfork bifurcation marking the onset of drop rotation is found to transition from supercritical to subcritical, giving rise to a field-strength interval of bistability. In this subcritical regime, the critical field is always large enough that, at the bifurcation, the symmetric base-state solution exhibits equatorial charge-density blowup singularities of the type recently described by Peng \textit{et al.} (Phys.~Rev.~Fluids, \textbf{9} 083701, 2024). As the rotation speed increases along the initially unstable solution branch from the bifurcation, the singularities gradually shift from the equator and ultimately disperse once the rotational component of the flow is strong enough to eliminate the surface stagnation points. 
\end{abstract}
\maketitle
%\newpage

\begin{figure}[t!]
\begin{center}
\includegraphics[scale=0.45,trim={1cm 0 1cm 0}]{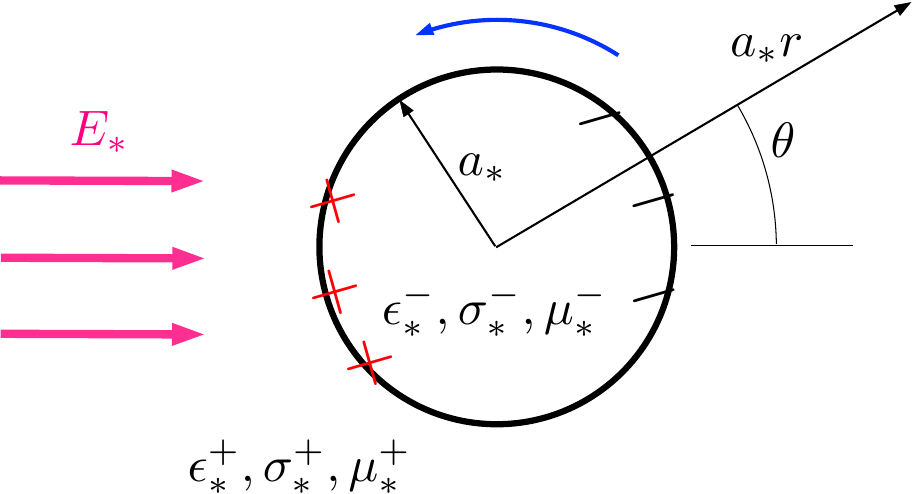}
\caption{Schematic of a circular drop undergoing Quincke rotation (or solid cylinder, corresponding to the case $\mu_*^-/\mu_*^+\to\infty$).}
\label{fig:sketch}
\end{center}
\end{figure}

\textit{Introduction.}---\citet{Quincke:1896} first observed that electrically insulating glass spheres and cylinders, as well as bubbles, spontaneously rotate when exposed to a sufficiently strong electric field, with the axis of rotation perpendicular to the field yet otherwise unconstrained. This striking symmetry-breaking phenomenon was elucidated  by \citet{Melcher:69}, who employed their leaky-dielectric electrohydrodynamic framework to develop a fully analytical model in the case of circular cylinders---later extended to spheres by \citet{Jones:84}. A key condition is that 
the charge-relaxation time---given by the ratio of permittivity to conductivity---is greater for the particle than for the suspending fluid. Under this condition, the charging of the particle boundary by bulk Ohmic currents results in it becoming polarized \emph{antiparallel} to the external field. A small misalignment of this induced surface-charge distribution generates an electric torque, driving rotation in a direction so as to amplify the perturbation through surface-charge convection (see Fig.~\ref{fig:sketch}). The onset of Quincke rotation is governed by the \emph{electric Reynolds number}, which quantifies the competition between this destabilizing convective mechanism and the realignment of the induced charge by bulk Ohmic currents. 

Contemporary studies of the Quincke effect have focused on electrorheological applications \cite{Lobry:99,Cebers:04,Pannacci:07,Pannacci:07a,Lemaire:08},  the self-propulsion and collective dynamics of Quincke rollers near boundaries 
\cite{Bricard:13,Das:13,Pradillo:19,Das:19,Belovs:20,Kokot:22,Das:23}, and the chaotic dynamics exhibited by inertial and non-isotropic particles
\cite{Cebers:00,Lemaire:02,Peters:05}. Beyond solid particles, significant interest has emerged in the Quincke rotation of liquid drops, with initial observations by \citet{Krause:98}, and \citet{Ha:00}, followed by more systematic investigations by \citet{Salipante:10,Salipante:13}. Compared to solid spheres, the onset of Quincke rotation was found at higher field strengths, with the critical field increasing as the drop viscosity is reduced relative to the suspending fluid. Furthermore, \citet{Salipante:10} found that sufficiently inviscid drops exhibit hysteresis: as the field strength increased, rotation commenced discontinuously at a critical field; upon decreasing the field, rotation ceased, also discontinuously, at a lower threshold. (For $4.6\,\mathrm{mm}$ castor oil drops in silicone oil, hysteresis was observed for drop-to-background viscosity ratios below $\approx 0.7$.) Incidentally, Quincke rotation has also been observed for particle-coated drops \cite{Ouriemi:15} and liquid filaments \cite{Raju:21}.

The Quincke rotation of highly viscous drops is well understood, with simulations \cite{Das:17a,Firouznia:23} and perturbations from solid-particle theory \cite{He:13,Das:17,Das:21} demonstrating good agreement with experiments. In contrast, the onset of Quincke rotation of lower-viscosity drops remains poorly understood, as numerical simulations fail at surprisingly weak fields due to the apparent formation of charge-density ``shocks'' about the drop equator \cite{Das:17,Firouznia:23}. 
This behavior was first noted by \citet{Lanauze:15}, who employed initial-value simulations to attempt continuation of Taylor's \cite{Taylor:66}  \emph{symmetric} (non-rotational) steady-state solution beyond the weak-field regime. In his pioneering study of drop electrohydrodynamics, Taylor showed that an electric field acting on its own induced surface charge drives electrohydrodynamic flow within and outside the drop, even under weak fields---unlike in the solid-particle case, where the fluid is quiescent prior to instability. For slowly relaxing drops polarized antiparallel to the electric field, this flow is directed from the poles to the equator, thus convecting oppositely signed charges towards that location. As explained by \citet{Lanauze:15}, with increasing field strength such convection can lead to steepening of the charge-density profile at the equator. Similar steepening effects have been observed in other electrohydrodynamic systems \cite{Firouznia:22}.

The formation of charge-density singularities in the symmetric steady state was recently elucidated by \citet{Peng:24}, within a two-dimensional (2D) leaky-dielectric model of a circular drop subjected to an external field. 
They showed via local analysis that the governing equations support a steady singularity structure where the charge density \emph{diverges} (rather than forming a shock, as previously hypothesized) with the $-1/3$ power of signed distance from the drop ``equator'' (i.e., the pair of surface points on the diameter normal to the external field). This blowup singularity describes the \textit{annihilation} of positive and negative charges at equal rates, convected towards the equator by a locally induced stagnation flow; globally, the rate of annihilation is balanced by Ohmic charging of the interface. By encoding this information into their numerical scheme, Peng \textit{et al}.~were able to continue the steady symmetric solution to arbitrary electric fields, surpassing the singularity threshold. 

Before Peng \textit{et al.}'s \cite{Peng:24} study, the circular-drop model was used to study Quincke rotation by \citet{Feng:02} and \citet{Yariv:16}. Feng's numerical calculations, however, were limited to moderately and highly viscous drops, while Yariv and Frankel's asymptotic analysis, conducted in the limit of large electric Reynolds, allowed for arbitrary viscosity ratios but addressed a regime far from the instability threshold. In this Letter, we investigate the electrohydrodynamic response of a circular drop without \textit{a priori} symmetry assumptions, allowing for arbitrary field strengths and viscosity ratios. This enables us to characterize the emergence of bistability in the onset of Quincke rotation. As we demonstrate, capturing this transition necessarily entails accounting for the possibility of charge-density singularities in the solutions.

\textit{Formulation.}---As illustrated in Fig.~\ref{fig:sketch}, we consider a circular drop of radius $a_*$, suspended in a density-matched background liquid and subjected to a uniform, constant external electric field of magnitude $E_*$. Both the drop and the suspending phase are modelled as leaky dielectrics, characterized by viscosities $\mu^{\pm}_*$, permittivities $\epsilon^{\pm}_*$ and conductivities $\sigma^{\pm}_*$. Here, the sign denotes whether the property pertains to the drop  ($-$) or the suspending fluid ($+$), while the asterisk indicates a dimensional quantity. The associated charge-relaxation time scales are denoted by $\tau^{\pm}_*=\epsilon^{\pm}_*/\sigma_*^{\pm}$. 
Inertia and deformation are neglected assuming sufficiently small scales and strong surface tension. Henceforth, we adopt a dimensionless convention where lengths are normalized by $a_*$, time by $\tau_*^+$, potentials by $a_*E_*$, surface-charge density by $\epsilon_*^+E_*$ and velocities by $v_*=\epsilon_*^+a_*E_*^2/\mu_*^+$. We shall formulate the problem in a non-rotating frame comoving with the drop, employing dimensionless polar coordinates $(r,\theta)$ originating at its center, with $r=1$ corresponding to the drop interface and $\theta=0$ aligned with the external field.   

For leaky dielectrics, the space charge carried by a fluid parcel decays exponentially on the charge-relaxation time scale. Accordingly, it is customary to assume that charge is confined to the interface, and we denote the surface-charge density by $q$. It follows that the electric potentials $\varphi^{\pm}$ in the drop and the suspending fluid satisfy Laplace's equation in their respective domains. At the interface, the potentials satisfy continuity, $\varphi^+=\varphi^-$, while Gauss's law gives 
\begin{equation}\label{Gauss}
q= \mathcal{S}\pd{\varphi^-}{r}-\pd{\varphi^+}{r},
\end{equation}
where $\mathcal{S}=\epsilon_*^-/\epsilon_*^+$ is the drop permittivity relative to the suspending fluid. Furthermore, the exterior potential satisfies the far-field condition that $\varphi^{+} \sim -r\cos\theta$ as $r\to\infty$, representing the approach to the external field. 

The velocity fields $\bu^{\pm}=u^{\pm}\be_r+v^{\pm}\be_{\theta}$ in the drop and suspending fluid satisfy the Stokes equations in their respective domains. 
At the interface, the velocity fields satisfy impermeability, $u^{\pm}=0$; 
continuity, $v^+=v^-$; and the tangential-stress balance
\begin{equation}\label{tstress}
\pd{v^+}{r}-v-\mathcal{M}\left(\pd{v^-}{r}-v\right)=q\pd{\varphi}{\theta},
\end{equation}
where $\mathcal{M}=\mu_*^-/\mu_*^+$ is the relative drop viscosity. (We omit the sign superscript when evaluating $v^{\pm}$ or $\varphi^{\pm}$ at the interface.)
Furthermore, the exterior velocity field satisfies the far-field condition 
\begin{equation}\label{flow far}
\bu^+ \to  -\bU \quad \text{as} \quad r\to\infty,
\end{equation}
where $\bU$ is the \textit{a priori} unknown velocity of the drop center in the laboratory frame.  

The charge density $q$ satisfies the interfacial charging equation
\begin{equation}\label{q eq}
\pd{q}{t}+\mathrm{Re}_E\pd{}{\theta}(qv)=\pd{\varphi^+}{r}-\mathcal{R}\pd{\varphi^{-}}{r},
\end{equation}
where $\mathcal{R}=\sigma_*^-/\sigma_*^+$ is the relative drop conductivity and $\mathrm{Re}_E={\epsilon_*^+}^2E_*^2/\mu_*^+\sigma_*^+$ 
is the electric Reynolds number based on the suspending phase, namely the relaxation time scale $\tau^{+}_*$ over the convection time scale $a_*/v_*$. The first and second terms on the left-hand side of \eqref{q eq} represent charge accumulation and convection, respectively, while the right-hand side represents charge relaxation  by bulk Ohmic currents. 

Integration of \eqref{q eq} over the interface shows that the net drop charge decays exponentially on the time scale $\tau^+_*$. It is thus natural to assume that the net charge vanishes at all times. In an unbounded fluid, this implies that the net electric force on the drop vanishes at all times---consequently, so does the net hydrodynamic force. In fact, this assumption was already implicit in the far-field condition \eqref{flow far}, since in 2D Stokes flow a net hydrodynamic force would require logarithmically growing velocities. It is the lack of such logarithmic growth that serves to determine the drop velocity $\bU$. In the SI, we show that the lab-frame drop velocity can be expressed as minus the surface average of the drop-frame fluid velocity: $\bU=-\langle v\be_{\theta}\rangle$.

Our interest is mainly in the the drop angular velocity, an output of the model defined as the surface-averaged azimuthal velocity component  (positive for counterclockwise rotation):
\begin{equation}\label{omega def}
\Omega=\langle v \rangle.
\end{equation}
We show in the SI that the hydrodynamic torque on the drop (normalized by $\mu_*^+a_*v_*$) is $-4\pi\Omega$, and that a balance with the electric torque on the drop yields the relation 
\begin{equation}\label{torque balance}
\Omega=-(1+\mathcal{S})^{-1}\langle q \sin\theta \rangle.
\end{equation}

Remarkably, the number of independent parameters can be reduced beyond dimensional analysis, via an argument generalizing that of \citet{Peng:24} to allow for asymmetric states exhibiting drop rotation. (However, we will only utilize this reduction when presenting numerical results.) Thus, we show in the SI that the electric potentials possess the reflection property $\partial\varphi^+/\partial r=-\partial\varphi^-/\partial r-2\cos\theta$ at $r=1$, whereby Gauss's law \eqref{Gauss} can be simplified to $q=(1+\mathcal{S})\partial\varphi^-/\partial r + 2\cos\theta$ and the charging equation \eqref{q eq} gives  
\begin{equation}\label{charging modified}
\pd{q}{t}+\mathrm{Re}_E\pd{}{\theta}(qv)+(1-\mathcal{C})q=-2\mathcal{C}\cos\theta,
\end{equation}
where we define the ``charging parameter'' $\mathcal{C}=(\mathcal{S}-\mathcal{R})/(1+\mathcal{S})<1$.  
The modified Gauss law and charging equation allow us to eliminate the exterior potential from the problem. Furthermore, the velocity fields satisfy the reflection property $\partial{v}^+/\partial{r}-v=-\left(\partial{v}^-/\partial{r}-v\right)-2\Omega$, whereby it follows from the stress balance \eqref{tstress} that the exterior flow can also be eliminated. Upon rescaling the interior potential to $\tilde{\varphi}^-=(1+\mathcal{S})\varphi^-$, the interior velocity to $\tilde{\bu}^-=(1+\mathcal{M})(1+\mathcal{S})\bu^-$ (similarly for $\Omega$ and $\bU$) and the electric Reynolds number to $\widetilde{\mathrm{Re}}=(1+\mathcal{M})^{-1}(1+\mathcal{S})^{-1}\mathrm{Re}_E$, the rescaled interior problem is found to be independent of $\mathcal{S}$, and, in the case where $\Omega$ vanishes, also independent of $\mathcal{M}$.   

\textit{Solid cylinder.}---We begin by examining the limit of a highly viscous drop ($\mathcal{M}\gg1$), where the drop flows as a rigid body, reducing the problem to the classical solid-cylinder scenario analyzed by \citet{Melcher:69}. In that scenario, the azimuthal velocity at the surface is uniform, $v\equiv\Omega$ [cf.~\eqref{omega def}], so that $\bU=-\Omega\langle\be_{\theta}\rangle=\bzero$. Furthermore, the modified charging equation \eqref{charging modified} is effectively linearized. At steady state, it simplifies  to a first-order ODE for $q(\theta)$, whose unique $2\pi$-periodic solution is
\begin{equation}\label{q profile}
q=\frac{2\mathcal{C}\left[(\mathcal{C}-1)\cos\theta-\Omega\mathrm{Re}_E\sin\theta\right]}{(1-\mathcal{C})^2+\Omega^2\mathrm{Re}_E^2}.
\end{equation} 
The torque balance \eqref{torque balance} then yields an algebraic equation for $\Omega$. The trivial solution, ${\Omega}=0$, always exists. It represents a symmetric base state where the drop and suspending fluid are both at rest, with the induced surface charge polarized parallel ($\mathcal{C}<0$) or antiparallel ($\mathcal{C}>0$) to the external field. In the antiparallel-polarization scenario, corresponding to the case where charge relaxation is slower in the drop phase, there exists a critical electric Reynolds number, $\mathrm{Re}_E\ub{c,\mathrm{sol}}$, beyond which there also exist solutions ${\Omega}=\pm {\Omega}\ub{\mathrm{sol}}$, where 
\refstepcounter{equation}
$$
\label{Om sol and Re c sol}
\Omega\ub{\mathrm{sol}}=\frac{1-\mathcal{C}}{\mathrm{Re}_E}\sqrt{\frac{{\mathrm{Re}_E}}{{{\mathrm{Re}_E}}\ub{c,\mathrm{sol}}}-1}, \quad \mathrm{Re}_E\ub{c,\mathrm{sol}}=\mathcal{C}^{-1}(1-\mathcal{C})^2(1+\mathcal{S}),
\eqno{(\theequation\mathrm{a},\mathrm{b})}
$$ 
These nontrivial solutions represent steady Quincke rotation in either the anticlockwise or clockwise direction. For later reference, the maximum angular speed follows from (\ref{Om sol and Re c sol}a): $\Omega\ub{\mathrm{sol}}_{\text{max}}=\tfrac{1}{2}\mathcal{C}(1-\mathcal{C})^{-1}(1+\mathcal{S})^{-1}$. 
A linear stability analysis shows that at the threshold $\mathrm{Re}_E=\mathrm{Re}_E\ub{c,\mathrm{sol}}$ the symmetric base state loses stability in favor of the Quincke states. Thus, Quincke rotation of a solid cylinder arises through a supercritical pitchfork bifurcation. 
\begin{figure*}[t!]
\begin{center}
\includegraphics[scale=1,trim={0cm 0 0cm 0}]{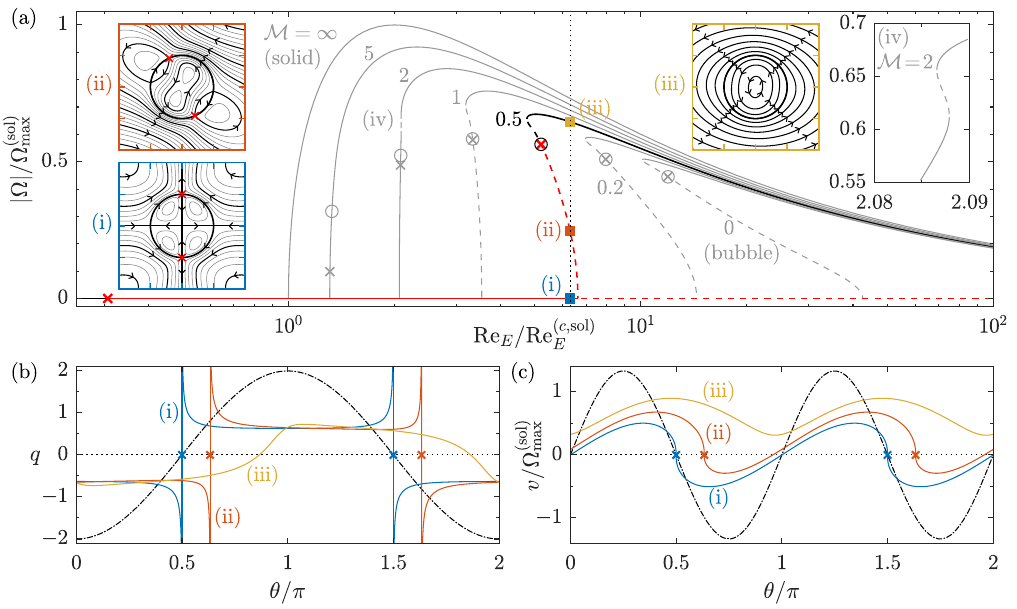}
\caption{Panel (a) shows steady angular speed as a function of $\mathrm{Re}_E$, respectively scaled by the maximum speed and critical $\mathrm{Re}_E$ for a solid cylinder, computed for $\mathcal{C}=0.5$ and several values of $\mathcal{M}$ as indicated. The rotational states exhibit surface stagnation points below the circles and charge-density blowup below the crosses. The case $\mathcal{M}=0.5$ is emphasized, with dashed segments indicating instability regions of the rotational and Taylor-symmetric states, and the cross along the horizontal axis indicating the singularity threshold of the Taylor-symmetric state. (For other values of $\mathcal{M}$ stability is indicated only for the rotational states.) We mark three states coexisting at the same $\mathrm{Re}_E$ and show their streamlines in the insets (i-iii) of panel (a); their charge-density profiles in panel (b); and their scaled surface-velocity profiles in panel (c). In panels (b,c), the crosses mark the locations of blowup singularities and the dash-dotted profiles represent the weak-field solution for $\mathrm{Re}_E=0$. The inset (iv) of panel (a) zooms in on the double-fold structure exhibited by the $\mathcal{M}=2$ rotational branch.}  \label{fig:bif}
\end{center}
\end{figure*}

\textit{Drops in weak fields}.---We now turn to drops having arbitrary viscosities, initially reviewing the weak-field regime $\mathrm{Re}_E\ll1$. In that limit, the electric and flow problems can be solved sequentially, furnishing the unique, globally attracting steady state in closed form \cite{Feng:02}. The electric problem is the same as for a solid cylinder under weak fields, or, more generally, a \emph{non-rotating} solid cylinder at arbitrary $\mathrm{Re}_E$. Hence, the charge density is obtained from \eqref{q profile} by setting $\Omega\mathrm{Re}_E=0$. In the drop scenario, however, the fluids are not stationary  under weak fields. Rather, the electrical shear stresses at the interface drive an electrohydrodynamic flow corresponding to the surface-velocity profile 
\begin{equation}
v=\frac{\mathcal{C}\sin 2\theta}{2(1-\mathcal{C})^2(1+\mathcal{M})(1+\mathcal{S})}.
\end{equation}

The above weak-field solution is the 2D analog of Taylor's approximation for spherical drops \cite{Taylor:66}. It is both fore-aft symmetric (i.e., about the equatorial line $\theta=\pm\pi/2$), such that $q$, $\varphi^{\pm}$ and $v^{\pm}$ are fore-aft antisymmetric and $u^{\pm}$ are fore-aft symmetric; and up-down symmetric (i.e., about the pole line $\theta=0,\pi$), such that $q$, $\varphi^{\pm}$ and $u^{\pm}$ are up-down symmetric and $v^{\pm}$ are up-down antisymmetric. This combined biaxial symmetry, henceforth referred to as ``Taylor symmetry,'' implies that the drop velocities $\bU$ and $\Omega$ both vanish. For $\mathcal{C}>0$, where the surface charge is polarized antiparallel to the external field, the surface velocity is directed from the poles to the equator; for $\mathcal{C}<0$, where the surface charge is polarized parallel to the external field, the surface velocity is directed from the equator to the poles. 

\textit{Charge-density blowup beyond weak fields}.---\citet{Peng:24} studied the continuation of the weak-field solution for a circular drop to arbitrary $\mathrm{Re}_E$. They found that this steady state preserved the Taylor symmetry of the weak-field solution, as well as the linkage between the polarization and  flow direction to the sign of the charging parameter $\mathcal{C}$. In the antiparallel-polarization scenario, where $\mathcal{C}>0$, they showed that past a critical $\mathrm{Re}_E$ the solution developed interfacial singularities at the equatorial points $\theta=\pm\pi/2$. Using local analysis, they discovered that the governing equations support steady charge-density blowup singularities of the form 
\begin{gather}
\label{local singular}
q \sim -\left(\frac{2J}{3}\right)^{1/3}(1+\mathcal{M})^{1/3}(1+\mathcal{S})^{1/3}
\Delta\theta^{-1/3}, \end{gather}
where $\Delta\theta=\theta\mp\pi/2$ and the constant $J=\lim_{\Delta\theta\to0}qv$ is the limiting value of the surface current. The singularity structure dictates a converging stagnation flow, however with the surface azimuthal velocity vanishing non-smoothly: $v\sim J/q$ as $\Delta\theta\to0$. Note that $|J|$ gives the rate of annihilation of positive and negative charges at the singularity. 
\begin{figure}[t!]
\begin{center}
\includegraphics[scale=0.45,trim={0.5cm 0 0 0}]{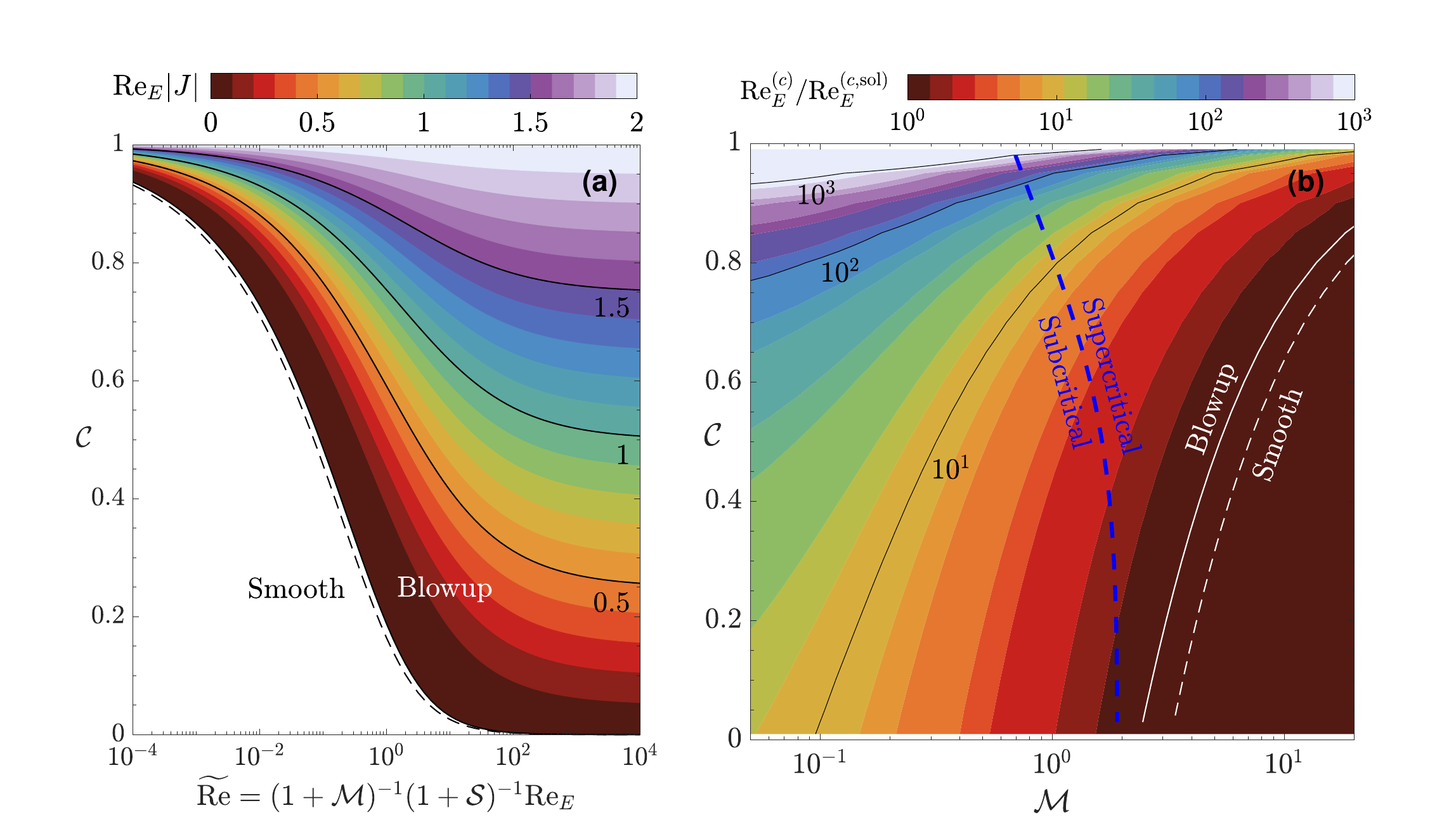}
\caption{(a) Charge-density blowup in the Taylor-symmetric state.  
(b) Onset of Quincke rotation.} 
\label{fig:maps}
\end{center}
\end{figure}

\citet{Peng:24} computed the Taylor-symmetric solution at arbitrary $\mathrm{Re}_E$ using a Fourier scheme where the singular behaviors \eqref{local singular} were subtracted, $J$ being calculated together with the solution using a global balance between Ohmic charging and annihilation. We here adopt a different approach which readily generalizes to non-symmetric solutions and is more accurate especially near the singularity threshold. As detailed in the SI, we use finite differences to solve an integral-equation formulation of the problem, with weak surface-charge diffusion (surface diffusivity $D_*$) included in order to regularize the charge-density singularities. The term $\mathrm{Pe}^{-1}\mathrm{Re}_E\partial^2{q}/\partial \theta^2$ is thus added to the right-hand-side of the charging balance \eqref{q eq}, where the P\'eclet number $\mathrm{Pe}=a_*v_*/D_*$. For $\mathrm{Pe}\gg1$, surface diffusion has a negligible effect except in narrow $O(\mathrm{Pe}^{-3/4})$ layers about singularities of the diffusionless problem. These layers smooth the charge-density profiles, with $q$ attaining large $O(\mathrm{Pe}^{1/4})$ values rather than diverging. In the simulations, we take $\mathrm{Pe}\,\mathrm{Re}_E^{-1}\gtrsim 10^{4}$ and resolve the narrow layers using an adapting exponentially stretched grid. 

The state labelled (i) in  Fig.~\ref{fig:bif} is an example of a Taylor-symmetric solution exhibiting charge-density blowup, computed for $\mathcal{C}=\mathcal{M}=0.5$ and $\widetilde{\mathrm{Re}}=2.1$ (corresponding to $\mathrm{Re}_E/\mathrm{Re}_E\ub{c,\mathrm{sol}}=6.3$ in this figure). The streamlines are shown in an inset in panel (a), while the charge-density and surface-velocity profiles are shown in panels (b,c) alongside the corresponding weak-field profiles. Fig.~\ref{fig:maps}(a) shows the region in the $\widetilde{\mathrm{Re}}\,$--$\,\mathcal{C}$ plane where the Taylor-symmetric state exhibits equatorial blowup, with the colour map depicting the singularity strength in that region. (As discussed by \citet{Peng:24}, there exists a narrow transition region between the regions of smooth and blowup solutions, where the charge density is continuous but nonsmooth at the equator.) 

\textit{Drop Quincke rotation}.---We now apply the regularized numerical scheme to investigate the emergence of spontaneous rotation in the circular-drop model. To this end, we employ a parameter-continuation method, beginning at large values of $\mathrm{Re}_E$ using the asymptotic theory of \citet{Yariv:16} developed in the limit $\mathrm{Re}_E\to\infty$. Their theory predicts steady rotational (and non-translational) states with angular velocity $\Omega \sim \pm\mathcal{C}^{1/2}(1+\mathcal{S})^{-1/2}\mathrm{Re}_E^{-1/2}\hat{\Omega}(\mathcal{M})$, 
where $\hat{\Omega}(\mathcal{M})$ is a numerical function of the viscosity ratio. This function decreases monotonically from $\hat{\Omega}\approx 1$ in the solid limit $\mathcal{M}\gg1$ [cf.~(\ref{Om sol and Re c sol}b)] to $\hat{\Omega}\approx 0.927$ in the bubble limit $\mathcal{M}\ll1$. The corresponding  charge-density profile describes an induced dipole oriented perpendicular to the external field, and the flow is characterized by closed streamlines around the drop similar to the Quincke states of a solid cylinder. Notably, in the absence of surface stagnation points, the charge-density profile remains smooth---contrasting the Taylor-symmetric state in this regime.

Fig.~\ref{fig:bif}(a) presents $|\Omega|/\Omega^{(\mathrm{sol})}_{\text{max}}$ values of computed steady states as a function of $\mathrm{Re}_E/\mathrm{Re}_E\ub{c,\mathrm{sol}}$ [cf.~\eqref{Om sol and Re c sol}], for $\mathcal{C}=0.5$ and several $\mathcal{M}$ values. Although the rotational states break the Taylor symmetry, they are found to retain a reduced ``$\pi$ symmetry'' compatible with the governing equations for all $\mathrm{Re}_E$, such that $q(\theta)=-q(\theta+\pi)$, ${\varphi}^{\pm}(r,\theta)=-{\varphi}^{\pm}(r,\theta+\pi)$, $u^\pm(r,\theta)=u^{\pm}(r,\theta+\pi)$ and ${v}^{\pm}(r,\theta)={v}^{\pm}(r,\theta+\pi)$. This reduced symmetry prohibits drop translation. 

We observe that the large-$\mathrm{Re}_E$ rotational solutions extend continuously down to the zero-angular-speed axis, i.e., the Taylor-symmetric solution. The critical electric Reynolds number $\mathrm{Re}_E=\mathrm{Re}_E\ub{c}$, where the rotational solutions bifurcate from the Taylor-symmetric solution, increases as the viscosity ratio $\mathcal{M}$ decreases between the solid and bubble limits. For sufficiently small $\mathcal{M}$ (below $\mathcal{M}\approx1.55$ for $\mathcal{C}=0.5$), the curve representing the rotational solutions exhibits a fold,
implying an interval of $\mathrm{Re}_E$ where four rotational solutions coexist with the Taylor-symmetric solution instead of just two. (Each rotational state with angular velocity $\Omega$ has a corresponding state with angular velocity $-\Omega$.) A numerical linear-stability analysis confirms that stability changes only at the bifurcation or at folds, with the Taylor-symmetric solution stable for sufficiently small $\mathrm{Re}_E$ and the rotational  solutions always stable for sufficiently large $\mathrm{Re}_E$. Thus, as $\mathcal{M}$ decreases, the pitchfork bifurcation where the rotational states emerge transitions from supercritical to \emph{subcritical}, leading to bistability between the larger-magnitude rotational states and the Taylor-symmetric state within the $\mathrm{Re}_E$ interval identified above. (For a range of $\mathcal{M}$ near this transition, the rotational branch actually folds twice, such that there is bistability between two pairs of rotational states in a narrow $\mathrm{Re}_E$ interval. This phenomenon is observed in inset (iv) of Fig.~\ref{fig:bif}(a) and in further detail in the SI.)

Since $\mathrm{Re}_E\ub{c}$ increases as $\mathcal{M}$ decreases, the Taylor-symmetric solution at the bifurcation may become singular for sufficiently small $\mathcal{M}$. 
 In that case, we find that charge-density blowup persists for the rotational states in a neighbourhood of the bifurcation, with the local singularity structure as in \eqref{local singular} but shifted from the drop equator in the direction of rotation. This behavior is illustrated in Fig.~\ref{fig:bif}(a), where rotational solutions are singular below the crosses, and exhibit surface stagnation points (necessary for singularity) below the circles. The cross on the zero-angular-speed axis marks the singularity threshold for the case $\mathcal{M}=0.5$, in which the onset of rotation is both singular and subcritical. For that value of $\mathcal{M}$, Fig.~\ref{fig:bif} presents  the streamlines, charge-density profiles and surface-velocity profiles corresponding to the different states coexisting at a $\mathrm{Re}_E$ value just below the instability threshold $\mathrm{Re}_E\ub{c}$: (i) a stable Taylor-symmetric state exhibiting equatorial blowup (discussed earlier); (ii) an unstable rotational state exhibiting a mix of open and closed streamlines due to competing rotational and straining electrohydrodynamic flows, with stagnation/singular points displaced from the equator; and (iii) a stable rotational state exhibiting conventional Quincke-like behavior characterized by closed streamlines and a smooth charge-density profile. 

In Fig.~\ref{fig:maps}(b), we present a phase map characterizing the onset of spontaneous rotation in the $\mathcal{M}\,$--$\,\mathcal{C}$ plane. The map distinguishes regions of smooth vs.~singular onset, and supercritical vs.~subcritical onset, with the color scale representing the normalized critical electric Reynolds number $\mathrm{Re}_E\ub{c}/\mathrm{Re}_E\ub{c,\mathrm{sol}}$. This map reveals that a subcritical onset is always singular, whereas a singular onset can be either supercritical or subcritical. This may explain why previous simulations not accounting for blowup singularities could not reproduce the bistability seen in experiments. 

Curiously, in certain parameter regimes ($\mathcal{C}\lesssim0.4$ and sufficiently small $\mathcal{M}$) our numerical investigations have uncovered steady-state solutions that break both the Taylor and $\pi$ symmetries---with the drop accordingly exhibiting both rotation and translation. These solutions, however, are 
\emph{unstable}, emerging from and reconnecting with continuously unstable rotational branches. A numerical demonstration of these intriguing states is provided in the SI. 

\textit{Concluding remarks.}---In the (``Taylor-symmetric'') base state of a circular drop polarized antiparallel to the electric field, electrohydrodynamic surface flow convects opposite charges toward the equator. This convection weakens the induced-charge dipole, as seen from the integral balance $\langle \br q\rangle = (1-\mathcal{C})^{-1}\left\{-\mathcal{C}\unit + \mathrm{Re}_E\langle qv \be_{\theta}\rangle\right\}$, derived in the SI, since $\unit\bcdot(qv \be_{\theta})>0$ under antiparallel polarization. The torque balance \eqref{torque balance} then suggests that convection reduces the rotational response to small perturbations that misalign the charge dipole, thus delaying the onset of Quincke rotation to higher field strengths. Away from the threshold, the Quincke state of a circular drop is qualitatively similar to that of a solid cylinder, so with decreasing field strength the Quincke branch can overshoot the postponed instability threshold, leading to a subcritical onset and bistability between the symmetric and Quincke states, consistently with experiments on lower-viscosity drops. As the rotation speed increases from the bifurcation, the drop gradually transitions from fluid-like to solid-like behavior. This transition is marked by the displacement of the stagnation points from the drop equator, in the direction of the rotation, and their eventual disappearance, accompanied by a transformation in streamline topology: from biaxial---straining-like outside the drop, cellular within the drop---to closed. 

The convection of opposite charges towards the drop equator may also lead to the formation of charge-density singularities, as described by \citet{Peng:24} for the Taylor-symmetric state. If these singularities are present at onset, they persist onto the Quincke branch in a neighbourhood of the bifurcation, albeit shifted from the equator along with the stagnation points, which are crucial for their existence. As the rate of rotation rate increases, the singularities disperse just before the surface stagnation points vanish. 

Both subcriticality and singularity emerge as the drop viscosity is reduced relative to the suspending fluid, corresponding to an enhanced effect of charge convection by the shear-driven electrohydrodynamic flow. A subcritical onset is always also singular, while a singular onset can be either subcritical or supercritical. Consequently, the subcritical regime can only be theoretically accessed by accounting for the possibility of charge-density singularities. This can be achieved through singularity-removal schemes, as used by \citet{Peng:24} for the symmetric state, or through  controlled regularization, as done here by incorporating weak surface-charge diffusion. An intriguing direction for future research is to explore how the charge-density singularities predicted by the leaky-dielectric equations are regularized \emph{physically}---potentially within an electrokinetic framework  \cite{Saville:97,Schnitzer:15,Mori:18,Ma:22}. 

In principle, Quincke rotation of a spherical drop could be studied using a similar approach, though the simulations would be more involved. We anticipate charge-density blowup occuring along curves---the equator in the symmetric state---with the same local singularity structure as in our present two-dimensional setup. However, in three dimensions, the combination of antiparallel polarization and strong convection could give rise to a competing surface-electroconvection instability of the type studied in the 1960s in meniscus and thin-film systems  \cite{Malkus:61,Melcher:69,Jolly:70}, leading to convection cells near the equator as observed for particle-coated drops \cite{Ouriemi:14,Ouriemi:15}.  This would entail studying the dynamics of such singular curves, and more generally the unsteady dynamics of charge-density singularities in the leaky-dielectric model. Lastly, incorporating deformation into our approach (either numerically or perturbatively) would be of obvious interest, especially since deformation is known to trigger secondary dynamical transitions under strong fields \cite{Salipante:13,He:13,Dong:23}. Notably, the normal stresses associated with charge-density blowup do not diverge rapidly enough to provoke singular deformations \cite{Peng:24}, indicating that the local singularity structure is robust. 

\textbf{Acknowledgements.} The authors acknowledge  support from the Leverhulme Trust through Research Project Grant RPG-2021-161.

\bibliography{refs}

%apsrev4-2.bst 2019-01-14 (MD) hand-edited version of apsrev4-1.bst
%Control: key (0)
%Control: author (8) initials jnrlst
%Control: editor formatted (1) identically to author
%Control: production of article title (0) allowed
%Control: page (0) single
%Control: year (1) truncated
%Control: production of eprint (0) enabled
\begin{thebibliography}{43}%
\makeatletter
\providecommand \@ifxundefined [1]{%
 \@ifx{#1\undefined}
}%
\providecommand \@ifnum [1]{%
 \ifnum #1\expandafter \@firstoftwo
 \else \expandafter \@secondoftwo
 \fi
}%
\providecommand \@ifx [1]{%
 \ifx #1\expandafter \@firstoftwo
 \else \expandafter \@secondoftwo
 \fi
}%
\providecommand \natexlab [1]{#1}%
\providecommand \enquote  [1]{``#1''}%
\providecommand \bibnamefont  [1]{#1}%
\providecommand \bibfnamefont [1]{#1}%
\providecommand \citenamefont [1]{#1}%
\providecommand \href@noop [0]{\@secondoftwo}%
\providecommand \href [0]{\begingroup \@sanitize@url \@href}%
\providecommand \@href[1]{\@@startlink{#1}\@@href}%
\providecommand \@@href[1]{\endgroup#1\@@endlink}%
\providecommand \@sanitize@url [0]{\catcode `\\12\catcode `\$12\catcode
  `\&12\catcode `\#12\catcode `\^12\catcode `\_12\catcode `\%12\relax}%
\providecommand \@@startlink[1]{}%
\providecommand \@@endlink[0]{}%
\providecommand \url  [0]{\begingroup\@sanitize@url \@url }%
\providecommand \@url [1]{\endgroup\@href {#1}{\urlprefix }}%
\providecommand \urlprefix  [0]{URL }%
\providecommand \Eprint [0]{\href }%
\providecommand \doibase [0]{https://doi.org/}%
\providecommand \selectlanguage [0]{\@gobble}%
\providecommand \bibinfo  [0]{\@secondoftwo}%
\providecommand \bibfield  [0]{\@secondoftwo}%
\providecommand \translation [1]{[#1]}%
\providecommand \BibitemOpen [0]{}%
\providecommand \bibitemStop [0]{}%
\providecommand \bibitemNoStop [0]{.\EOS\space}%
\providecommand \EOS [0]{\spacefactor3000\relax}%
\providecommand \BibitemShut  [1]{\csname bibitem#1\endcsname}%
\let\auto@bib@innerbib\@empty
%</preamble>
\bibitem [{\citenamefont {Quincke}(1896)}]{Quincke:1896}%
  \BibitemOpen
  \bibfield  {author} {\bibinfo {author} {\bibfnamefont {G.}~\bibnamefont
  {Quincke}},\ }\bibfield  {title} {\bibinfo {title} {Ueber rotationen im
  constanten electrischen felde},\ }\href@noop {} {\bibfield  {journal}
  {\bibinfo  {journal} {Ann. Phys. Chem.}\ }\textbf {\bibinfo {volume} {Band
  59}},\ \bibinfo {pages} {417} (\bibinfo {year} {1896})}\BibitemShut {NoStop}%
\bibitem [{\citenamefont {Melcher}\ and\ \citenamefont
  {Taylor}(1969)}]{Melcher:69}%
  \BibitemOpen
  \bibfield  {author} {\bibinfo {author} {\bibfnamefont {J.~R.}\ \bibnamefont
  {Melcher}}\ and\ \bibinfo {author} {\bibfnamefont {G.~I.}\ \bibnamefont
  {Taylor}},\ }\bibfield  {title} {\bibinfo {title} {Electrohydrodynamics: a
  review of the role of interfacial shear stresses},\ }\href@noop {} {\bibfield
   {journal} {\bibinfo  {journal} {Ann. Rev. Fluid Mech.}\ }\textbf {\bibinfo
  {volume} {1}},\ \bibinfo {pages} {111} (\bibinfo {year} {1969})}\BibitemShut
  {NoStop}%
\bibitem [{\citenamefont {Jones}(1984)}]{Jones:84}%
  \BibitemOpen
  \bibfield  {author} {\bibinfo {author} {\bibfnamefont {T.~B.}\ \bibnamefont
  {Jones}},\ }\bibfield  {title} {\bibinfo {title} {Quincke rotation of
  spheres},\ }\href@noop {} {\bibfield  {journal} {\bibinfo  {journal} {IEEE
  Trans. Ind. Appl.}\ ,\ \bibinfo {pages} {845}} (\bibinfo {year}
  {1984})}\BibitemShut {NoStop}%
\bibitem [{\citenamefont {Lobry}\ and\ \citenamefont
  {Lemaire}(1999)}]{Lobry:99}%
  \BibitemOpen
  \bibfield  {author} {\bibinfo {author} {\bibfnamefont {L.}~\bibnamefont
  {Lobry}}\ and\ \bibinfo {author} {\bibfnamefont {E.}~\bibnamefont
  {Lemaire}},\ }\bibfield  {title} {\bibinfo {title} {Viscosity decrease
  induced by a dc electric field in a suspension},\ }\href@noop {} {\bibfield
  {journal} {\bibinfo  {journal} {J. Electrostat.}\ }\textbf {\bibinfo {volume}
  {47}},\ \bibinfo {pages} {61} (\bibinfo {year} {1999})}\BibitemShut {NoStop}%
\bibitem [{\citenamefont {C{\=e}bers}(2004)}]{Cebers:04}%
  \BibitemOpen
  \bibfield  {author} {\bibinfo {author} {\bibfnamefont {A.}~\bibnamefont
  {C{\=e}bers}},\ }\bibfield  {title} {\bibinfo {title} {Bistability and
  “negative” viscosity for a suspension of insulating particles in an
  electric field},\ }\href@noop {} {\bibfield  {journal} {\bibinfo  {journal}
  {Phys. Rev. Lett.}\ }\textbf {\bibinfo {volume} {92}},\ \bibinfo {pages}
  {034501} (\bibinfo {year} {2004})}\BibitemShut {NoStop}%
\bibitem [{\citenamefont {Pannacci}\ \emph
  {et~al.}(2007{\natexlab{a}})\citenamefont {Pannacci}, \citenamefont
  {Lemaire},\ and\ \citenamefont {Lobry}}]{Pannacci:07}%
  \BibitemOpen
  \bibfield  {author} {\bibinfo {author} {\bibfnamefont {N.}~\bibnamefont
  {Pannacci}}, \bibinfo {author} {\bibfnamefont {E.}~\bibnamefont {Lemaire}},\
  and\ \bibinfo {author} {\bibfnamefont {L.}~\bibnamefont {Lobry}},\ }\bibfield
   {title} {\bibinfo {title} {Rheology and structure of a suspension of
  particles subjected to quincke rotation},\ }\href@noop {} {\bibfield
  {journal} {\bibinfo  {journal} {Rheol. acta}\ }\textbf {\bibinfo {volume}
  {46}},\ \bibinfo {pages} {899} (\bibinfo {year}
  {2007}{\natexlab{a}})}\BibitemShut {NoStop}%
\bibitem [{\citenamefont {Pannacci}\ \emph
  {et~al.}(2007{\natexlab{b}})\citenamefont {Pannacci}, \citenamefont {Lobry},\
  and\ \citenamefont {Lemaire}}]{Pannacci:07a}%
  \BibitemOpen
  \bibfield  {author} {\bibinfo {author} {\bibfnamefont {N.}~\bibnamefont
  {Pannacci}}, \bibinfo {author} {\bibfnamefont {L.}~\bibnamefont {Lobry}},\
  and\ \bibinfo {author} {\bibfnamefont {E.}~\bibnamefont {Lemaire}},\
  }\bibfield  {title} {\bibinfo {title} {How insulating particles increase the
  conductivity of a suspension},\ }\href@noop {} {\bibfield  {journal}
  {\bibinfo  {journal} {Phys. Rev. Lett.}\ }\textbf {\bibinfo {volume} {99}},\
  \bibinfo {pages} {094503} (\bibinfo {year} {2007}{\natexlab{b}})}\BibitemShut
  {NoStop}%
\bibitem [{\citenamefont {Lemaire}\ \emph {et~al.}(2008)\citenamefont
  {Lemaire}, \citenamefont {Lobry}, \citenamefont {Pannacci},\ and\
  \citenamefont {Peters}}]{Lemaire:08}%
  \BibitemOpen
  \bibfield  {author} {\bibinfo {author} {\bibfnamefont {E.}~\bibnamefont
  {Lemaire}}, \bibinfo {author} {\bibfnamefont {L.}~\bibnamefont {Lobry}},
  \bibinfo {author} {\bibfnamefont {N.}~\bibnamefont {Pannacci}},\ and\
  \bibinfo {author} {\bibfnamefont {F.}~\bibnamefont {Peters}},\ }\bibfield
  {title} {\bibinfo {title} {Viscosity of an electro-rheological suspension
  with internal rotations},\ }\href@noop {} {\bibfield  {journal} {\bibinfo
  {journal} {J. Rheol.}\ }\textbf {\bibinfo {volume} {52}},\ \bibinfo {pages}
  {769} (\bibinfo {year} {2008})}\BibitemShut {NoStop}%
\bibitem [{\citenamefont {Bricard}\ \emph {et~al.}(2013)\citenamefont
  {Bricard}, \citenamefont {Caussin}, \citenamefont {Desreumaux}, \citenamefont
  {Dauchot},\ and\ \citenamefont {Bartolo}}]{Bricard:13}%
  \BibitemOpen
  \bibfield  {author} {\bibinfo {author} {\bibfnamefont {A.}~\bibnamefont
  {Bricard}}, \bibinfo {author} {\bibfnamefont {J.}~\bibnamefont {Caussin}},
  \bibinfo {author} {\bibfnamefont {N.}~\bibnamefont {Desreumaux}}, \bibinfo
  {author} {\bibfnamefont {O.}~\bibnamefont {Dauchot}},\ and\ \bibinfo {author}
  {\bibfnamefont {D.}~\bibnamefont {Bartolo}},\ }\bibfield  {title} {\bibinfo
  {title} {Emergence of macroscopic directed motion in populations of motile
  colloids},\ }\href@noop {} {\bibfield  {journal} {\bibinfo  {journal}
  {Nature}\ }\textbf {\bibinfo {volume} {503}},\ \bibinfo {pages} {95}
  (\bibinfo {year} {2013})}\BibitemShut {NoStop}%
\bibitem [{\citenamefont {Das}\ and\ \citenamefont
  {Saintillan}(2013)}]{Das:13}%
  \BibitemOpen
  \bibfield  {author} {\bibinfo {author} {\bibfnamefont {D.}~\bibnamefont
  {Das}}\ and\ \bibinfo {author} {\bibfnamefont {D.}~\bibnamefont
  {Saintillan}},\ }\bibfield  {title} {\bibinfo {title} {Electrohydrodynamic
  interaction of spherical particles under quincke rotation},\ }\href@noop {}
  {\bibfield  {journal} {\bibinfo  {journal} {Phys. Rev. E}\ }\textbf {\bibinfo
  {volume} {87}},\ \bibinfo {pages} {043014} (\bibinfo {year}
  {2013})}\BibitemShut {NoStop}%
\bibitem [{\citenamefont {Pradillo}\ \emph {et~al.}(2019)\citenamefont
  {Pradillo}, \citenamefont {Karani},\ and\ \citenamefont
  {Vlahovska}}]{Pradillo:19}%
  \BibitemOpen
  \bibfield  {author} {\bibinfo {author} {\bibfnamefont {G.~E.}\ \bibnamefont
  {Pradillo}}, \bibinfo {author} {\bibfnamefont {H.}~\bibnamefont {Karani}},\
  and\ \bibinfo {author} {\bibfnamefont {P.~M.}\ \bibnamefont {Vlahovska}},\
  }\bibfield  {title} {\bibinfo {title} {Quincke rotor dynamics in confinement:
  rolling and hovering},\ }\href@noop {} {\bibfield  {journal} {\bibinfo
  {journal} {Soft matter}\ }\textbf {\bibinfo {volume} {15}},\ \bibinfo {pages}
  {6564} (\bibinfo {year} {2019})}\BibitemShut {NoStop}%
\bibitem [{\citenamefont {Das}\ and\ \citenamefont {Lauga}(2019)}]{Das:19}%
  \BibitemOpen
  \bibfield  {author} {\bibinfo {author} {\bibfnamefont {D.}~\bibnamefont
  {Das}}\ and\ \bibinfo {author} {\bibfnamefont {E.}~\bibnamefont {Lauga}},\
  }\bibfield  {title} {\bibinfo {title} {Active particles powered by quincke
  rotation in a bulk fluid},\ }\href@noop {} {\bibfield  {journal} {\bibinfo
  {journal} {Phys. Rev. Lett.}\ }\textbf {\bibinfo {volume} {122}},\ \bibinfo
  {pages} {194503} (\bibinfo {year} {2019})}\BibitemShut {NoStop}%
\bibitem [{\citenamefont {Belovs}\ and\ \citenamefont
  {C{\=e}bers}(2020)}]{Belovs:20}%
  \BibitemOpen
  \bibfield  {author} {\bibinfo {author} {\bibfnamefont {M.}~\bibnamefont
  {Belovs}}\ and\ \bibinfo {author} {\bibfnamefont {A.}~\bibnamefont
  {C{\=e}bers}},\ }\bibfield  {title} {\bibinfo {title} {Quincke rotation
  driven flows},\ }\href@noop {} {\bibfield  {journal} {\bibinfo  {journal}
  {Phys. Rev. Fluids}\ }\textbf {\bibinfo {volume} {5}},\ \bibinfo {pages}
  {013701} (\bibinfo {year} {2020})}\BibitemShut {NoStop}%
\bibitem [{\citenamefont {Kokot}\ \emph {et~al.}(2022)\citenamefont {Kokot},
  \citenamefont {Faizi}, \citenamefont {Pradillo}, \citenamefont {Snezhko},\
  and\ \citenamefont {Vlahovska}}]{Kokot:22}%
  \BibitemOpen
  \bibfield  {author} {\bibinfo {author} {\bibfnamefont {G.}~\bibnamefont
  {Kokot}}, \bibinfo {author} {\bibfnamefont {H.~A.}\ \bibnamefont {Faizi}},
  \bibinfo {author} {\bibfnamefont {G.~E.}\ \bibnamefont {Pradillo}}, \bibinfo
  {author} {\bibfnamefont {A.}~\bibnamefont {Snezhko}},\ and\ \bibinfo {author}
  {\bibfnamefont {P.~M.}\ \bibnamefont {Vlahovska}},\ }\bibfield  {title}
  {\bibinfo {title} {Spontaneous self-propulsion and nonequilibrium shape
  fluctuations of a droplet enclosing active particles},\ }\href@noop {}
  {\bibfield  {journal} {\bibinfo  {journal} {Comm. Phys.}\ }\textbf {\bibinfo
  {volume} {5}},\ \bibinfo {pages} {91} (\bibinfo {year} {2022})}\BibitemShut
  {NoStop}%
\bibitem [{\citenamefont {Das}\ and\ \citenamefont
  {Saintillan}(2023)}]{Das:23}%
  \BibitemOpen
  \bibfield  {author} {\bibinfo {author} {\bibfnamefont {D.}~\bibnamefont
  {Das}}\ and\ \bibinfo {author} {\bibfnamefont {D.}~\bibnamefont
  {Saintillan}},\ }\bibfield  {title} {\bibinfo {title} {On the absence of
  collective motion in a bulk suspension of spontaneously rotating dielectric
  particles},\ }\href@noop {} {\bibfield  {journal} {\bibinfo  {journal} {Soft
  Matter}\ }\textbf {\bibinfo {volume} {19}},\ \bibinfo {pages} {6825}
  (\bibinfo {year} {2023})}\BibitemShut {NoStop}%
\bibitem [{\citenamefont {C{\=e}bers}\ \emph {et~al.}(2000)\citenamefont
  {C{\=e}bers}, \citenamefont {Lemaire},\ and\ \citenamefont
  {Lobry}}]{Cebers:00}%
  \BibitemOpen
  \bibfield  {author} {\bibinfo {author} {\bibfnamefont {A.}~\bibnamefont
  {C{\=e}bers}}, \bibinfo {author} {\bibfnamefont {E.}~\bibnamefont
  {Lemaire}},\ and\ \bibinfo {author} {\bibfnamefont {L.}~\bibnamefont
  {Lobry}},\ }\bibfield  {title} {\bibinfo {title} {Electrohydrodynamic
  instabilities and orientation of dielectric ellipsoids in low-conducting
  fluids},\ }\href@noop {} {\bibfield  {journal} {\bibinfo  {journal} {Phys.
  Rev. E}\ }\textbf {\bibinfo {volume} {63}},\ \bibinfo {pages} {016301}
  (\bibinfo {year} {2000})}\BibitemShut {NoStop}%
\bibitem [{\citenamefont {Lemaire}\ and\ \citenamefont
  {Lobry}(2002)}]{Lemaire:02}%
  \BibitemOpen
  \bibfield  {author} {\bibinfo {author} {\bibfnamefont {E.}~\bibnamefont
  {Lemaire}}\ and\ \bibinfo {author} {\bibfnamefont {L.}~\bibnamefont
  {Lobry}},\ }\bibfield  {title} {\bibinfo {title} {Chaotic behavior in
  electro-rotation},\ }\href@noop {} {\bibfield  {journal} {\bibinfo  {journal}
  {Physica A Stat.}\ }\textbf {\bibinfo {volume} {314}},\ \bibinfo {pages}
  {663} (\bibinfo {year} {2002})}\BibitemShut {NoStop}%
\bibitem [{\citenamefont {Peters}\ \emph {et~al.}(2005)\citenamefont {Peters},
  \citenamefont {Lobry},\ and\ \citenamefont {Lemaire}}]{Peters:05}%
  \BibitemOpen
  \bibfield  {author} {\bibinfo {author} {\bibfnamefont {F.}~\bibnamefont
  {Peters}}, \bibinfo {author} {\bibfnamefont {L.}~\bibnamefont {Lobry}},\ and\
  \bibinfo {author} {\bibfnamefont {E.}~\bibnamefont {Lemaire}},\ }\bibfield
  {title} {\bibinfo {title} {Experimental observation of lorenz chaos in the
  quincke rotor dynamics},\ }\href@noop {} {\bibfield  {journal} {\bibinfo
  {journal} {Chaos}\ }\textbf {\bibinfo {volume} {15}} (\bibinfo {year}
  {2005})}\BibitemShut {NoStop}%
\bibitem [{\citenamefont {Krause}\ and\ \citenamefont
  {Chandratreya}(1998)}]{Krause:98}%
  \BibitemOpen
  \bibfield  {author} {\bibinfo {author} {\bibfnamefont {S.}~\bibnamefont
  {Krause}}\ and\ \bibinfo {author} {\bibfnamefont {P.}~\bibnamefont
  {Chandratreya}},\ }\bibfield  {title} {\bibinfo {title} {Electrorotation of
  deformable fluid droplets},\ }\href@noop {} {\bibfield  {journal} {\bibinfo
  {journal} {J. Colloid Interface Sci.}\ }\textbf {\bibinfo {volume} {206}},\
  \bibinfo {pages} {10} (\bibinfo {year} {1998})}\BibitemShut {NoStop}%
\bibitem [{\citenamefont {Ha}\ and\ \citenamefont {Yang}(2000)}]{Ha:00}%
  \BibitemOpen
  \bibfield  {author} {\bibinfo {author} {\bibfnamefont {J.-W.}\ \bibnamefont
  {Ha}}\ and\ \bibinfo {author} {\bibfnamefont {S.-M.}\ \bibnamefont {Yang}},\
  }\bibfield  {title} {\bibinfo {title} {Electrohydrodynamics and
  electrorotation of a drop with fluid less conductive than that of the ambient
  fluid},\ }\href@noop {} {\bibfield  {journal} {\bibinfo  {journal} {Phys.
  Fluids}\ }\textbf {\bibinfo {volume} {12}},\ \bibinfo {pages} {764} (\bibinfo
  {year} {2000})}\BibitemShut {NoStop}%
\bibitem [{\citenamefont {Salipante}\ and\ \citenamefont
  {Vlahovska}(2010)}]{Salipante:10}%
  \BibitemOpen
  \bibfield  {author} {\bibinfo {author} {\bibfnamefont {P.~F.}\ \bibnamefont
  {Salipante}}\ and\ \bibinfo {author} {\bibfnamefont {P.~M.}\ \bibnamefont
  {Vlahovska}},\ }\bibfield  {title} {\bibinfo {title} {Electrohydrodynamics of
  drops in strong uniform {DC} electric fields},\ }\href@noop {} {\bibfield
  {journal} {\bibinfo  {journal} {Phys. Fluids}\ }\textbf {\bibinfo {volume}
  {22}},\ \bibinfo {pages} {112110} (\bibinfo {year} {2010})}\BibitemShut
  {NoStop}%
\bibitem [{\citenamefont {Salipante}\ and\ \citenamefont
  {Vlahovska}(2013)}]{Salipante:13}%
  \BibitemOpen
  \bibfield  {author} {\bibinfo {author} {\bibfnamefont {P.~F.}\ \bibnamefont
  {Salipante}}\ and\ \bibinfo {author} {\bibfnamefont {P.~M.}\ \bibnamefont
  {Vlahovska}},\ }\bibfield  {title} {\bibinfo {title} {Electrohydrodynamic
  rotations of a viscous droplet},\ }\href@noop {} {\bibfield  {journal}
  {\bibinfo  {journal} {Phys. Rev. E}\ }\textbf {\bibinfo {volume} {88}},\
  \bibinfo {pages} {043003} (\bibinfo {year} {2013})}\BibitemShut {NoStop}%
\bibitem [{\citenamefont {Ouriemi}\ and\ \citenamefont
  {Vlahovska}(2015)}]{Ouriemi:15}%
  \BibitemOpen
  \bibfield  {author} {\bibinfo {author} {\bibfnamefont {M.}~\bibnamefont
  {Ouriemi}}\ and\ \bibinfo {author} {\bibfnamefont {P.~M.}\ \bibnamefont
  {Vlahovska}},\ }\bibfield  {title} {\bibinfo {title} {Electrohydrodynamic
  deformation and rotation of a particle-coated drop},\ }\href@noop {}
  {\bibfield  {journal} {\bibinfo  {journal} {Langmuir}\ }\textbf {\bibinfo
  {volume} {31}},\ \bibinfo {pages} {6298} (\bibinfo {year}
  {2015})}\BibitemShut {NoStop}%
\bibitem [{\citenamefont {Raju}\ \emph {et~al.}(2021)\citenamefont {Raju},
  \citenamefont {Kyriakopoulos},\ and\ \citenamefont {Timonen}}]{Raju:21}%
  \BibitemOpen
  \bibfield  {author} {\bibinfo {author} {\bibfnamefont {G.}~\bibnamefont
  {Raju}}, \bibinfo {author} {\bibfnamefont {N.}~\bibnamefont
  {Kyriakopoulos}},\ and\ \bibinfo {author} {\bibfnamefont {J.~V.~I.}\
  \bibnamefont {Timonen}},\ }\bibfield  {title} {\bibinfo {title} {Diversity of
  non-equilibrium patterns and emergence of activity in confined
  electrohydrodynamically driven liquids},\ }\href@noop {} {\bibfield
  {journal} {\bibinfo  {journal} {Sci. Adv.}\ }\textbf {\bibinfo {volume}
  {7}},\ \bibinfo {pages} {eabh1642} (\bibinfo {year} {2021})}\BibitemShut
  {NoStop}%
\bibitem [{\citenamefont {Das}\ and\ \citenamefont
  {Saintillan}(2017{\natexlab{a}})}]{Das:17a}%
  \BibitemOpen
  \bibfield  {author} {\bibinfo {author} {\bibfnamefont {D.}~\bibnamefont
  {Das}}\ and\ \bibinfo {author} {\bibfnamefont {D.}~\bibnamefont
  {Saintillan}},\ }\bibfield  {title} {\bibinfo {title} {Electrohydrodynamics
  of viscous drops in strong electric fields: numerical simulations},\
  }\href@noop {} {\bibfield  {journal} {\bibinfo  {journal} {J. Fluid Mech.}\
  }\textbf {\bibinfo {volume} {829}},\ \bibinfo {pages} {127} (\bibinfo {year}
  {2017}{\natexlab{a}})}\BibitemShut {NoStop}%
\bibitem [{\citenamefont {Firouznia}\ \emph {et~al.}(2023)\citenamefont
  {Firouznia}, \citenamefont {Bryngelson},\ and\ \citenamefont
  {Saintillan}}]{Firouznia:23}%
  \BibitemOpen
  \bibfield  {author} {\bibinfo {author} {\bibfnamefont {M.}~\bibnamefont
  {Firouznia}}, \bibinfo {author} {\bibfnamefont {S.~H.}\ \bibnamefont
  {Bryngelson}},\ and\ \bibinfo {author} {\bibfnamefont {D.}~\bibnamefont
  {Saintillan}},\ }\bibfield  {title} {\bibinfo {title} {A spectral boundary
  integral method for simulating electrohydrodynamic flows in viscous drops},\
  }\href@noop {} {\bibfield  {journal} {\bibinfo  {journal} {J. Comp. Phys.}\
  ,\ \bibinfo {pages} {112248}} (\bibinfo {year} {2023})}\BibitemShut {NoStop}%
\bibitem [{\citenamefont {He}\ \emph {et~al.}(2013)\citenamefont {He},
  \citenamefont {Salipante},\ and\ \citenamefont {Vlahovska}}]{He:13}%
  \BibitemOpen
  \bibfield  {author} {\bibinfo {author} {\bibfnamefont {H.}~\bibnamefont
  {He}}, \bibinfo {author} {\bibfnamefont {P.~F.}\ \bibnamefont {Salipante}},\
  and\ \bibinfo {author} {\bibfnamefont {P.~M.}\ \bibnamefont {Vlahovska}},\
  }\bibfield  {title} {\bibinfo {title} {Electrorotation of a viscous droplet
  in a uniform direct current electric field},\ }\href@noop {} {\bibfield
  {journal} {\bibinfo  {journal} {Phys. Fluids}\ }\textbf {\bibinfo {volume}
  {25}} (\bibinfo {year} {2013})}\BibitemShut {NoStop}%
\bibitem [{\citenamefont {Das}\ and\ \citenamefont
  {Saintillan}(2017{\natexlab{b}})}]{Das:17}%
  \BibitemOpen
  \bibfield  {author} {\bibinfo {author} {\bibfnamefont {D.}~\bibnamefont
  {Das}}\ and\ \bibinfo {author} {\bibfnamefont {D.}~\bibnamefont
  {Saintillan}},\ }\bibfield  {title} {\bibinfo {title} {A nonlinear
  small-deformation theory for transient droplet electrohydrodynamics},\
  }\href@noop {} {\bibfield  {journal} {\bibinfo  {journal} {J. Fluid Mech.}\
  }\textbf {\bibinfo {volume} {810}},\ \bibinfo {pages} {225} (\bibinfo {year}
  {2017}{\natexlab{b}})}\BibitemShut {NoStop}%
\bibitem [{\citenamefont {Das}\ and\ \citenamefont
  {Saintillan}(2021)}]{Das:21}%
  \BibitemOpen
  \bibfield  {author} {\bibinfo {author} {\bibfnamefont {D.}~\bibnamefont
  {Das}}\ and\ \bibinfo {author} {\bibfnamefont {D.}~\bibnamefont
  {Saintillan}},\ }\bibfield  {title} {\bibinfo {title} {A three-dimensional
  small-deformation theory for electrohydrodynamics of dielectric drops},\
  }\href@noop {} {\bibfield  {journal} {\bibinfo  {journal} {J. Fluid Mech.}\
  }\textbf {\bibinfo {volume} {914}},\ \bibinfo {pages} {A22} (\bibinfo {year}
  {2021})}\BibitemShut {NoStop}%
\bibitem [{\citenamefont {Lanauze}\ \emph {et~al.}(2015)\citenamefont
  {Lanauze}, \citenamefont {Walker},\ and\ \citenamefont {Khair}}]{Lanauze:15}%
  \BibitemOpen
  \bibfield  {author} {\bibinfo {author} {\bibfnamefont {J.~A.}\ \bibnamefont
  {Lanauze}}, \bibinfo {author} {\bibfnamefont {L.~M.}\ \bibnamefont
  {Walker}},\ and\ \bibinfo {author} {\bibfnamefont {A.~S.}\ \bibnamefont
  {Khair}},\ }\bibfield  {title} {\bibinfo {title} {Nonlinear
  electrohydrodynamics of slightly deformed oblate drops},\ }\href@noop {}
  {\bibfield  {journal} {\bibinfo  {journal} {J. Fluid Mech.}\ }\textbf
  {\bibinfo {volume} {774}},\ \bibinfo {pages} {245} (\bibinfo {year}
  {2015})}\BibitemShut {NoStop}%
\bibitem [{\citenamefont {Taylor}(1966)}]{Taylor:66}%
  \BibitemOpen
  \bibfield  {author} {\bibinfo {author} {\bibfnamefont {G.}~\bibnamefont
  {Taylor}},\ }\bibfield  {title} {\bibinfo {title} {{Studies in
  electrohydrodynamics. I. The circulation produced in a drop by electrical
  field}},\ }\href@noop {} {\bibfield  {journal} {\bibinfo  {journal} {Proc.
  Roy. Soc. London A}\ }\textbf {\bibinfo {volume} {291}},\ \bibinfo {pages}
  {159} (\bibinfo {year} {1966})}\BibitemShut {NoStop}%
\bibitem [{\citenamefont {Firouznia}\ \emph {et~al.}(2022)\citenamefont
  {Firouznia}, \citenamefont {Miksis}, \citenamefont {Vlahovska},\ and\
  \citenamefont {Saintillan}}]{Firouznia:22}%
  \BibitemOpen
  \bibfield  {author} {\bibinfo {author} {\bibfnamefont {M.}~\bibnamefont
  {Firouznia}}, \bibinfo {author} {\bibfnamefont {M.~J.}\ \bibnamefont
  {Miksis}}, \bibinfo {author} {\bibfnamefont {P.~M.}\ \bibnamefont
  {Vlahovska}},\ and\ \bibinfo {author} {\bibfnamefont {D.}~\bibnamefont
  {Saintillan}},\ }\bibfield  {title} {\bibinfo {title} {Instability of a
  planar fluid interface under a tangential electric field in a stagnation
  point flow},\ }\href@noop {} {\bibfield  {journal} {\bibinfo  {journal} {J.
  Fluid Mech.}\ }\textbf {\bibinfo {volume} {931}},\ \bibinfo {pages} {A25}
  (\bibinfo {year} {2022})}\BibitemShut {NoStop}%
\bibitem [{\citenamefont {Peng}\ \emph {et~al.}(2024)\citenamefont {Peng},
  \citenamefont {Brandao}, \citenamefont {Yariv},\ and\ \citenamefont
  {Schnitzer}}]{Peng:24}%
  \BibitemOpen
  \bibfield  {author} {\bibinfo {author} {\bibfnamefont {G.~G.}\ \bibnamefont
  {Peng}}, \bibinfo {author} {\bibfnamefont {R.}~\bibnamefont {Brandao}},
  \bibinfo {author} {\bibfnamefont {E.}~\bibnamefont {Yariv}},\ and\ \bibinfo
  {author} {\bibfnamefont {O.}~\bibnamefont {Schnitzer}},\ }\bibfield  {title}
  {\bibinfo {title} {Equatorial blowup and polar caps in drop
  electroohydrodynamics},\ }\href@noop {} {\bibfield  {journal} {\bibinfo
  {journal} {Phys. Rev. Fluids}\ } (\bibinfo {year} {2024})}\BibitemShut
  {NoStop}%
\bibitem [{\citenamefont {Feng}(2002)}]{Feng:02}%
  \BibitemOpen
  \bibfield  {author} {\bibinfo {author} {\bibfnamefont {J.~Q.}\ \bibnamefont
  {Feng}},\ }\bibfield  {title} {\bibinfo {title} {A {2D} electrohydrodynamic
  model for electrorotation of fluid drops},\ }\href@noop {} {\bibfield
  {journal} {\bibinfo  {journal} {J. Colloid Interface Sci.}\ }\textbf
  {\bibinfo {volume} {246}},\ \bibinfo {pages} {112} (\bibinfo {year}
  {2002})}\BibitemShut {NoStop}%
\bibitem [{\citenamefont {Yariv}\ and\ \citenamefont
  {Frankel}(2016)}]{Yariv:16}%
  \BibitemOpen
  \bibfield  {author} {\bibinfo {author} {\bibfnamefont {E.}~\bibnamefont
  {Yariv}}\ and\ \bibinfo {author} {\bibfnamefont {I.}~\bibnamefont
  {Frankel}},\ }\bibfield  {title} {\bibinfo {title} {Electrohydrodynamic
  rotation of drops at large electric reynolds numbers},\ }\href@noop {}
  {\bibfield  {journal} {\bibinfo  {journal} {J. Fluid Mech.}\ }\textbf
  {\bibinfo {volume} {788}},\ \bibinfo {pages} {R2} (\bibinfo {year}
  {2016})}\BibitemShut {NoStop}%
\bibitem [{\citenamefont {Saville}(1997)}]{Saville:97}%
  \BibitemOpen
  \bibfield  {author} {\bibinfo {author} {\bibfnamefont {D.~A.}\ \bibnamefont
  {Saville}},\ }\bibfield  {title} {\bibinfo {title} {Electrohydrodynamics: the
  {Taylor--Melcher} leaky dielectric model},\ }\href@noop {} {\bibfield
  {journal} {\bibinfo  {journal} {Ann. Rev. Fluid Mech.}\ }\textbf {\bibinfo
  {volume} {29}},\ \bibinfo {pages} {27} (\bibinfo {year} {1997})}\BibitemShut
  {NoStop}%
\bibitem [{\citenamefont {Schnitzer}\ and\ \citenamefont
  {Yariv}(2015)}]{Schnitzer:15}%
  \BibitemOpen
  \bibfield  {author} {\bibinfo {author} {\bibfnamefont {O.}~\bibnamefont
  {Schnitzer}}\ and\ \bibinfo {author} {\bibfnamefont {E.}~\bibnamefont
  {Yariv}},\ }\bibfield  {title} {\bibinfo {title} {The {T}aylor--{M}elcher
  leaky dielectric model as a macroscale electrokinetic description},\
  }\href@noop {} {\bibfield  {journal} {\bibinfo  {journal} {J. Fluid Mech.}\
  }\textbf {\bibinfo {volume} {773}},\ \bibinfo {pages} {1} (\bibinfo {year}
  {2015})}\BibitemShut {NoStop}%
\bibitem [{\citenamefont {Mori}\ and\ \citenamefont {Young}(2018)}]{Mori:18}%
  \BibitemOpen
  \bibfield  {author} {\bibinfo {author} {\bibfnamefont {Y.}~\bibnamefont
  {Mori}}\ and\ \bibinfo {author} {\bibfnamefont {Y.-N.}\ \bibnamefont
  {Young}},\ }\bibfield  {title} {\bibinfo {title} {From electrodiffusion
  theory to the electrohydrodynamics of leaky dielectrics through the weak
  electrolyte limit},\ }\href@noop {} {\bibfield  {journal} {\bibinfo
  {journal} {J. Fluid Mech.}\ }\textbf {\bibinfo {volume} {855}},\ \bibinfo
  {pages} {67} (\bibinfo {year} {2018})}\BibitemShut {NoStop}%
\bibitem [{\citenamefont {Ma}\ \emph {et~al.}(2022)\citenamefont {Ma},
  \citenamefont {Booty},\ and\ \citenamefont {Siegel}}]{Ma:22}%
  \BibitemOpen
  \bibfield  {author} {\bibinfo {author} {\bibfnamefont {M.}~\bibnamefont
  {Ma}}, \bibinfo {author} {\bibfnamefont {M.~R.}\ \bibnamefont {Booty}},\ and\
  \bibinfo {author} {\bibfnamefont {M.}~\bibnamefont {Siegel}},\ }\bibfield
  {title} {\bibinfo {title} {A model for the electric field-driven flow and
  deformation of a drop or vesicle in strong electrolyte solutions},\
  }\href@noop {} {\bibfield  {journal} {\bibinfo  {journal} {J. Fluid Mech.}\
  }\textbf {\bibinfo {volume} {943}},\ \bibinfo {pages} {A47} (\bibinfo {year}
  {2022})}\BibitemShut {NoStop}%
\bibitem [{\citenamefont {Malkus}\ and\ \citenamefont
  {Veronis}(1961)}]{Malkus:61}%
  \BibitemOpen
  \bibfield  {author} {\bibinfo {author} {\bibfnamefont {W.~V.~R.}\
  \bibnamefont {Malkus}}\ and\ \bibinfo {author} {\bibfnamefont
  {G.}~\bibnamefont {Veronis}},\ }\bibfield  {title} {\bibinfo {title} {Surface
  electroconvection},\ }\href@noop {} {\bibfield  {journal} {\bibinfo
  {journal} {Phys. Fluids}\ }\textbf {\bibinfo {volume} {4}},\ \bibinfo {pages}
  {13} (\bibinfo {year} {1961})}\BibitemShut {NoStop}%
\bibitem [{\citenamefont {Jolly}\ and\ \citenamefont
  {Melcher}(1970)}]{Jolly:70}%
  \BibitemOpen
  \bibfield  {author} {\bibinfo {author} {\bibfnamefont {D.~C.}\ \bibnamefont
  {Jolly}}\ and\ \bibinfo {author} {\bibfnamefont {J.~R.}\ \bibnamefont
  {Melcher}},\ }\bibfield  {title} {\bibinfo {title} {Electroconvective
  instability in a fluid layer},\ }\href@noop {} {\bibfield  {journal}
  {\bibinfo  {journal} {Proc. Roy. Soc. London. A}\ }\textbf {\bibinfo {volume}
  {314}},\ \bibinfo {pages} {269} (\bibinfo {year} {1970})}\BibitemShut
  {NoStop}%
\bibitem [{\citenamefont {Ouriemi}\ and\ \citenamefont
  {Vlahovska}(2014)}]{Ouriemi:14}%
  \BibitemOpen
  \bibfield  {author} {\bibinfo {author} {\bibfnamefont {M.}~\bibnamefont
  {Ouriemi}}\ and\ \bibinfo {author} {\bibfnamefont {P.~M.}\ \bibnamefont
  {Vlahovska}},\ }\bibfield  {title} {\bibinfo {title} {Electrohydrodynamics of
  particle-covered drops},\ }\href@noop {} {\bibfield  {journal} {\bibinfo
  {journal} {J. Fluid Mech.}\ }\textbf {\bibinfo {volume} {751}},\ \bibinfo
  {pages} {106} (\bibinfo {year} {2014})}\BibitemShut {NoStop}%
\bibitem [{\citenamefont {Dong}\ and\ \citenamefont {Sau}(2023)}]{Dong:23}%
  \BibitemOpen
  \bibfield  {author} {\bibinfo {author} {\bibfnamefont {Q.}~\bibnamefont
  {Dong}}\ and\ \bibinfo {author} {\bibfnamefont {A.}~\bibnamefont {Sau}},\
  }\bibfield  {title} {\bibinfo {title} {Unsteady electrorotation of a viscous
  drop in a uniform electric field},\ }\href@noop {} {\bibfield  {journal}
  {\bibinfo  {journal} {Phys. Fluids}\ }\textbf {\bibinfo {volume} {35}}
  (\bibinfo {year} {2023})}\BibitemShut {NoStop}%
\end{thebibliography}%
\end{document}